# Evaluating Feature Selection Methods for Macro-Economic Forecasting, Applied for Iran's macro-economic variables

**Mahdi Goldani**

m.goldani@hsu.ac.ir

***Abstract:***

*This paper reviews the different methods of feature selection applied to macroeconomics for the development indicators of Iran using World Bank indicators. This paper will, therefore, focus on the comprehensive evaluation framework consisting of RMSE and MAE within a 10-fold cross-validation setup and make a systematic ranking in respect to different methodologies of feature selection. Besides, the study has distinctly stated that Stepwise Selection, Tree-based methods, Hausdorff distance, Euclidean distance, and Mutual Information Score contributed much toward the predictive errors by reducing them and thus being more efficient than others. Other methods, such as Recursive Feature Elimination with Cross-Validation (RFECV) and Variance Thresholding, were relatively less effective. The results underline the robustness of the similarity-based approaches, especially Hausdorff and Euclidean distances, since their performance on the various datasets tested was very good, with an average rank of 9.125 in the range of methods used. Therefore, these results will become critical in the use of different types of feature selection methods with significant implications for the improvement of predictive accuracy for models used in economic analysis and planning. The finding recommends that researchers and practitioners who deal with complex economic datasets should focus on stepwise and tree-based methods along with similarity-based techniques.*

## Introduction

The challenge of high-dimensional data lies in the exponential increase in complexity and sparsity it introduces. Furthermore, the storage and transmission costs escalate, visualization becomes intricate, and feature redundancy or irrelevance often plagues analysis [1]. Addressing these challenges necessitates the deployment of dimensionality reduction techniques, feature selection, and regularization methods, along with meticulous data preprocessing, to distill relevant insights while mitigating the adverse impacts of high dimensionality on machine learning and data analysis tasks. Feature selection is one of the techniques used for dimensionality reduction. Feature selection involves carefully choosing a subset of significant features (variables or predictors) for model creation. This pivotal step constitutes an integral component of the broader data preprocessing procedure. Within dimensionality reduction techniques, feature selection emerges as a prominent strategy. This approach entails the identification and retention of pertinent features while simultaneously eliminating any irrelevant or redundant ones [2]. Feature selection is a critical step in the machine learning and data analysis process, particularly when dealing with high-dimensional datasets. By identifying and selecting the most relevant features, feature selection enhances model performance by improving predictive accuracy, reducing overfitting, and lowering computational costs. In the context of these target variables, the feature selection process plays a crucial role in obtaining more accurate forecasts. By systematically identifying and retaining the most relevant features while eliminating redundant or irrelevant ones, feature selection enhances the model's ability to capture the underlying patterns and relationships within the data. This leads to improved predictive accuracy, as measured by lower RMSE and MAE values. For example, accurately forecasting GDP growth or inflation rates requires isolating key economic indicators that directly influence these outcomes, such as broad money supply, government expenditure, and foreign direct investment. Similarly, understanding the drivers of unemployment or manufacturing growth necessitates focusing on the most impactful predictors. Effective feature selection ensures that the models are not only more accurate but also more interpretable and computationally efficient, facilitating better economic analysis and policymaking. By reducing the complexity of the models, feature selection also helps prevent overfitting, ensuring that the forecasts remain robust and reliable when applied to new data.

Feature selection is the process of selection best and appropriate subsets of features which improve and enhance robustness of forecasting. This process takes place in preprocessing steps. Before the any training and testing in machine learning methods, it’s necessary to select the most reverent features base on target value. Beside all FS methods that are presented in literature, there are some methods that can detect the most relevant feature such as time series similarity methods. Review the literature show there is no paper which use similarity methods as FS methods. But there are some in common between these two methods that make them good alternative. Measuring similarity in time series forms the basis for the clustering and classification of these data, and its task is to measure the distance between two time series. The smaller the distance between the target variable and a feature, the feature can be included as a related feature in the model, and therefore a subset of features are considered for the model that have the smallest distance with the target variable. Therefore, the main question of this research is whether time series similarity methods work as well as feature selection methods in selecting a subset of features? The answer this question is because the simplicity of the pre-processing step is as important as the efficiency of the methods used and it helps to save time.

Feature selection is a widely used technique in various data mining and machine learning application. In literature of feature selection there is no study that use similarity methods directly as feature selection methods but there are some researches explore this concept or incorporate similarity measures into feature selection processes. For example, Zhu et al [3] In the proposed Feature Selection-based Feature Clustering (FSFC) algorithm, similarity-based feature clustering utilized a means of unsupervised feature selection. Mitra [4] propose an unsupervised feature selection algorithm designed for large datasets with high dimensionality. The algorithm is focused on measuring the similarity between features to identify and remove redundancy, resulting in a more efficient and effective feature selection process. In the domain of software defect prediction, Yu et al. [5] emphasize the central role of similarity in gauging the likeness or proximity among distinct software modules (referred to as samples) based on their respective features. Shi et al. [6] proposed a novel approach called Adaptive-Similarity-based Multi-modality Feature Selection (ASMFS) for multimodal classification in Alzheimer's disease (AD) and its prodromal stage, mild cognitive impairment (MCI). They addressed the limitations of traditional methods, which often rely on pre-defined similarity matrices to depict data structure, making it challenging to accurately capture the intrinsic relationships across different modalities in high-dimensional space. In the FU's [7] article Following the evaluation of feature relevance, redundant features are identified and removed using feature similarity. Features that exhibit high similarity to one another are considered redundant and are consequently eliminated from the dataset. Feature similarity measures are utilized to quantify the similarity between pairs of features. These measures help identify redundant features by assessing their degree of resemblance or closeness.

The literature review indicates that time series similarity methods are not used as an independent method in feature selection, so as a new idea, the performance of time series similarity methods to select a subset of features with performance Feature selection methods are compared. The rest of the paper is organized as follows: methodology is discussed in Section 2, Section 3 presents results of the study, Section 4 reports discussion on findings and conclusions.

## Methodology

The methodology illustrated in Figure 4 was followed for the viral incidence time series forecasting, where the following five steps were systematically applied:

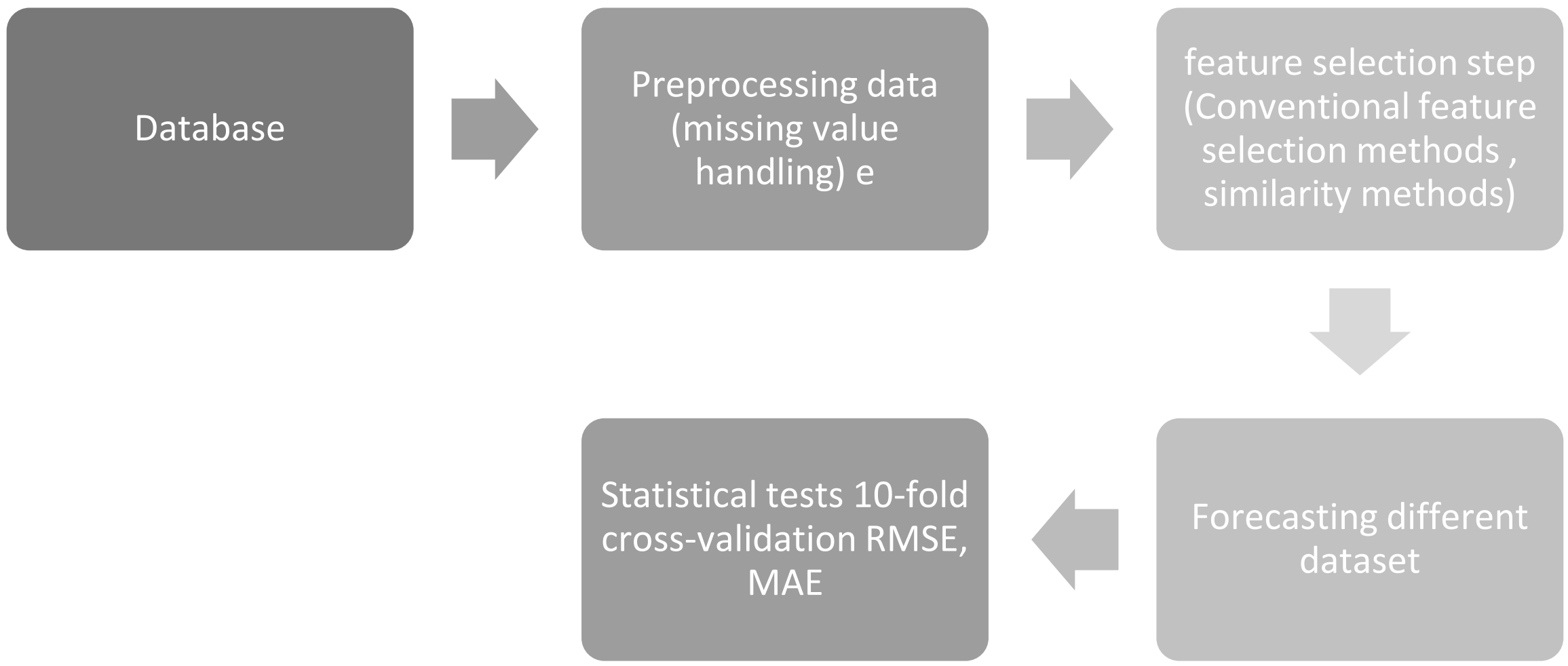

Figure 4. The complete methodology

**Dataset**

The aim of this paper compares the prediction performance of a subset of data selected by feature selection methods and time series similarity methods. The dataset chosen to achieve this goal is World Bank Development Indicators. To compare and confidence of efficiency of each dataset chosen based on the FS method, the methods were implemented on the World Bank dataset with different target variables as follows; Adjusted savings consumption of fixed capital, Broad money, Food production index (2014-2016 = 100), Foreign direct investment, net inflows (% of GDP), gdp growth, General government final consumption expenditure (% of GDP), GNI, Gross domestic income, Gross domestic saving, Gross national expenditure (% of GDP), Gross value added at basic prices, Households and NPISHs Final consumption expenditure per capita (constant 2015 US$), Imports of goods and services (constant 2015 US$), Manufacturing, value added (annual % growth), Official exchange rate (LCU per US$, period average), Stocks traded, total value (% of GDP), Total debt service (% of exports of goods, services and primary income), Unemployment, total (% of total labor force) (modeled ILO estimate), Wholesale price index (2010 = 100) and Inflation consumer prices. According to all target values we had 20 different datasets and we used FS methods to choose the best subset of this dataset. The data of this article was extracted from the World Bank website from 1990 to 2022.

**Preprocessing data (missing value handling)**

The first step of this study is Data Preparation. The World Development Indicators dataset from the World Bank is employed for analysis. Focused on Iran, the dataset spans from 1960 to 2022. In this study, the inflation variable is treated as the target, while the remaining variables are handled as independent variables. As a first step, before starting to choose features, we remove the variables with more than 80% missing data. For the remaining variables with missing data less than 80%, we use the K-Nearest Neighbors (KNN) method to fill in the gaps.

**feature selection methods**

Once the database without missing value is obtained, the next step is to apply FS to each of them. Since the similarity methods are classified as Filtered methods, in this research, filtered methods are examined as standard feature selection methods. Filtered feature selection methods are techniques that select features based on their relationship with the target variable, as opposed to evaluating the relationships between features themselves. These methods typically rely on statistical tests to determine the relevance of features for the predictive model. Examples of filtered feature selection methods include techniques like, correlation coefficients, and information gain, among others. These methods are efficient in high-dimensional datasets and are often used as a preliminary step in feature selection before employing more complex algorithms. The most well-known flirted methods are correlation-based, variance threshold, and information gain.

Table2. Filtered feature selection methods

| | **Method Name** | **Definition** | **Disadvantage** |
|---|---|---|---|
| Filters methods | Correlation-based | Identifies the strength and direction of the linear relationship between two variables. | Assumes only linear relationships and may miss out on nonlinear associations. |
| | Variance Threshold | Eliminates features with low variance, considering them less informative. | It cannot capture nonlinear relationships and might overlook useful features with low variance. |
| | Information Gain | Measures the effectiveness of a feature in classifying the data, often used in decision trees. | They might struggle with continuous data, and selecting features based on information gain may not be sufficient for complex datasets. |
| Wrappers methods | Forward Selection | It is a greedy algorithm that starts with an empty set of features and iteratively adds one feature at a time based on certain criteria, such as improving the performance of the model. | Overfitting is a concern, and it is sensitive to the initial set of features, potentially missing global patterns. Despite limitations, it is practical for simplicity in resource-constrained scenarios, requiring careful consideration of criteria for feature selection. |
| | Backward Elimination | Backward Elimination is a feature selection method that starts with all features and iteratively removes the least significant ones based on a chosen criterion, often improving the model's performance at each step. | One disadvantage of Backward Elimination is that it does not allow for the addition of features in later steps, limiting its ability to reconsider decisions and potentially resulting in a suboptimal feature subset. |
| | Recursive Feature Elimination | Recursive Feature Elimination is a feature selection technique that systematically removes the least important features from the model, typically by recursively training the model and assessing feature importance until the desired number of features is reached. It helps identify the most relevant subset of features for optimal model performance. | One potential disadvantage of RFE is its computational intensity, especially when dealing with a large number of features, as it involves repeatedly training the model and evaluating feature importance. Additionally, it may not perform well in cases where features interact in complex ways or when the relationship between features and the target variable is non-linear. |
| | Stepwise Selection | Stepwise Selection is a feature selection method that involves iteratively adding or removing features from a model based on certain criteria. There are two main types: Forward Selection and Backward Elimination. In Forward Selection, features are added one at a time, while in Backward Elimination, features are removed iteratively. The process continues until a predefined criterion, such as model performance or a significance level, is met. | A potential disadvantage of Stepwise Selection is its sensitivity to the order in which features are added or removed, which can lead to suboptimal subsets. Additionally, the stepwise nature may not consider interactions between features effectively, and the final subset chosen may be influenced by the stopping criterion selected. Careful consideration of criteria and potential overfitting is essential when applying stepwise selection methods. |
| | Genetic Algorithms | Genetic Algorithms are optimization methods inspired by natural selection. They iteratively evolve a population of potential solutions, applying genetic operators like crossover and mutation. Fitness evaluation guides the selection of individuals for the next generation. Despite their effectiveness, GAs can be computationally intensive, and tuning parameters is crucial. | A drawback is their computational complexity, especially for large search spaces, and the challenge of parameter tuning. GAs may not guarantee finding the global optimum and are sensitive to parameter choices and problem characteristics. |
| | Simulated Annealing | Simulated Annealing is a probabilistic optimization algorithm inspired by annealing in metallurgy. It explores solutions by allowing both uphill and downhill movements, preventing it from getting stuck in local optima. However, its effectiveness depends on parameter choices, and convergence rates may vary. | Sensitivity to parameter choices, such as the cooling schedule, and variable convergence rates depending on the problem make Simulated Annealing less efficient for certain optimization tasks. |
| Embedded methods | L1 Regularization (Lasso) | Lasso is a regularization method in machine learning that adds a penalty term to the model's cost function, promoting sparsity in the coefficients and performing feature selection. | Sensitivity to the choice of the regularization parameter ($\lambda$) is a potential drawback, requiring careful tuning for optimal results. |
| | Tree-based methods | Tree-based methods, like Random Forest and Gradient Boosted Trees, use decision trees to capture complex patterns in data through recursive feature splits. | Prone to overfitting, especially with deep trees, requiring regularization techniques and parameter tuning for better generalization. |
| | Recursive Feature Elimination with Cross-Validation (RFECV) | RFECV combines Recursive Feature Elimination (RFE) and cross-validation to iteratively select an optimal subset of features and evaluate model performance. | The method can be computationally intensive due to multiple cross-validation iterations and may have limitations with non-linear relationships and complex feature interactions. |
| | XGBoost and LightGBM | XGBoost and LightGBM are powerful gradient boosting frameworks in machine learning. They efficiently build decision tree ensembles, but proper hyperparameter tuning is crucial to prevent overfitting. | Both XGBoost and LightGBM can be sensitive to hyperparameter tuning, and improper settings may lead to overfitting. Careful parameter selection is essential for optimal performance. |

## The proposed approach

The proposed approach is classified as a Filter method that measures the relevance of features by their correlation with the dependent variable. Fig5 shows the framework of the proposed approach which consists of four main steps:

Fig5. The framework of the proposed feature selection

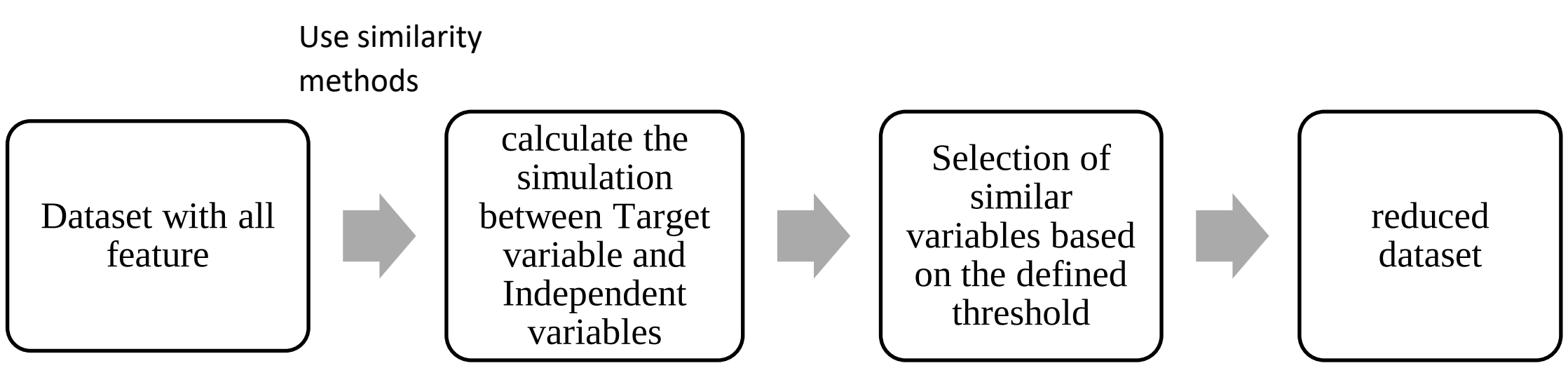

The primary step and algorithm in this approach is similarity methods. In this study, we evaluate FS with several different distance measures. The distance measures used are the Euclidean distance measure, Dynamic time warping, Edit Distance on Real sequence, Longest Common Subsequence, and Edit Distance with Real Penalty distance measure.

## Similarity methods

Measuring similarity in time series forms the basis for the clustering and classification of these data, and its task is to measure the distance between two time series. Similarity in time series plays a vital role in analysing temporal patterns. Firstly, the similarity between time series has been used as an absolute measure for statistical inference about the relationship between time series from different data sets [25]. In recent years, the increase in data collection has made it possible to create time series data. In the past few years, tasks such as regression, classification, clustering, and segmentation were employed for working with time series. In many cases, these tasks require defining a distance measurement that indicates the level of similarity between time series. Therefore, studying various methods for measuring the distance between time series appears essential and necessary. Among the different types of similarity measurement criteria for time series, they can be divided into three categories: step-by-step measures, distribution-based measures, and geometric methods.

Step-by-step measures include Euclidean distance and correlation. Euclidean distance is the first distance function used as a similarity measurement criterion for time series. It is notable for its computational simplicity, although it applies to sub paths of equal length and does not consider local time shifts. The correlation coefficient is a method for estimating the degree of correlation between two-time series. It ranges from -1 to 1, and it is unsurprising that the Pearson correlation coefficient is closely related to the Euclidean distance among normal distributions.

Distribution-based measures include Dynamic Time Warping (DTW) (unlike Euclidean distance, in this method, each point of the first time series can be compared with any arbitrary point of the second time series, provided that their differences are minimized), Longest Common Subsequence (LCSS) (this method is a variant of edit distance and its main idea is to allow some unique elements by allowing the stretching of two series without rearranging the order of elements), Edit Distance on Real Sequences (EDR) (EDR is based on edit distance over strings and by quantizing the distance between a pair of elements to 0 and 1, it eliminates noise effects), and Edit Distance with Real Penalty (ERP) (by combining L1-norm and edit distance, ERP can support local time shift and uses L1-norm between two non-gap elements), and Time Warp Edit Distance (TWED) (a metric similarity measure that allows flexibility even in terms of time axis alignment).

Geometric measures include Hausdorff (the minimum supremum distance from a point in set 1 to any other point in set 2), Fréchet distance (it is the minimum leash length required to connect a dog and its owner), and SSPD (calculates the point's distance from line segments for all sample points of the reference path and all line segments of the other path, and then reports the average distance).

Table 3. Similarity methods

| Method | Advantages | Disadvantages |
|---|---|---|
| Euclidean distance | The most straightforward, clearest, and most widely used criteria<br><br>No need for parameter estimation | The exact length of two-time series,<br><br>Lack of local time shift support.<br><br>Inefficiency with increasing dimensions of the time series.<br><br>The sensitivity of Euclidean distance to small changes in the time axis |
| DTW) Dynamic Time Warping( | The DTW interval performs local scaling for the time dimension and ensures the preservation of the order of the time series samples. Any point of the first time series can be compared with any arbitrary point of the second series, provided that their differences are minimized. One of the advantages of this distance function is its ability to measure the distance between time series with different lengths and support the local time | Being time-consuming<br><br>Sensitivity to noise, the heavy computational load required to find the optimal time-alignment path, incorrect clustering due to a large amount of outliers at the beginning and end of the sequence (some elements may not be comparable where DTW must find all elements match.) |
| | shift. | limiting the time deviation,<br><br>The need to calculate some costly Lp norms,<br><br>Not being metric and<br><br>Need to pair all elements in a series |
| LCSS) Longest Common SubSequence( | It is robust against noise and, in addition to giving more weight to similar parts of the series, provides an intuitive concept between paths. By focusing on the common parts, it gives the correct clustering. It enables more efficient approximate calculations. In this method, unlike the Euclidean distance, the data do not need to be normalized. | The results of time series data mining under LCSS strongly depend on the similarity threshold because the similarity measurement approach in LCSS is a zero-and-one approach. Since there is no information about the data and it is tough to determine the correct similarity threshold, using LCSS can lead to poor results. It is not metric and does not obey the triangle inequality. |
| EDR) Edit Distance on Real sequence( | Existing distance functions are usually sensitive to noise, change, and data scaling, which usually occurs due to sensor failure, errors in detection techniques, disturbance signals, and different sampling rates. It is not always possible to wipe the data to remove these items. EDR is robust against data corruption. | not metric<br><br>The reason that it does not obey the triangle inequality is that when a gap is created, it repeats the previous element. |
| ERP) Edit Distance with Real Penalty( | ERP is the only distance metric regardless of the Lp norm used, but it works better for regular series, especially for determining the gap g value.<br><br>ERP is a method based on editing distance that benefits from the advantages of DTW and EDR. This criterion uses a reference point to measure the time gap between matching samples. ERP transforms EDR into a metric whose distance function follows the triangle inequality law. | Because this method involves time thresholding, two locations will not be compared if the difference between their time indexes is too large. |
| TWED )Time Wrap Edit Distance( | Includes DTW and LCSS features. This method controls the time warp as a coefficient for the deviation penalty in the time dimension. | The originality of TWED, compared to ERP, is apart from managing the addition and deletion of hard parameter introductions. The classification error rate is very sensitive to the hardness parameter, while it shows some regular behavior. |
| Hasdorf | Hausdorff distance is a metric measure. It measures the distance between two sets of metric spaces.<br><br>It shows the spatial similarity between two routes and measures how far they are from each other. | When two curves have a small Hausdorff distance but are not generally similar, in this case, the Hausdorff distance is not suitable. The reason for this disagreement is that the Hausdorff distance only considers the set of points of both curves and does not reflect the trend of the curves. However, the trend is vital in many applications, such as handwriting recognition. |
| | | Hausdorff distance, in addition to route samples, all points in between<br><br>It also considers samples, which complicates the calculation of this<br><br>becomes the standard.<br><br>They have been widely used in many domains where shape comparison is needed, but they cannot generally compare paths. The Frechet and Hausdorff distance returns the maximum distance between two objects at given points in the two objects. |
| Fraishe is discrete | It considers samples and their order in a continuous sequence. Iter et al. used the discrete Freiche distance based on the regression model to estimate the continuous Freiche distance<br><br>developed This method reduces the complexity of Freishe's criterion | They have been widely used in many domains where shape comparison is needed, but they cannot generally compare paths. The Frechet and Hausdorff distance returns the maximum distance between two objects at given points in the two objects. |
| SSPD )Symmetric Segment Path Distance( | Because the SSPD method is the sum of the Euclidean distances and considering that it is based on point to segment-,, it has solved the problem of the Euclidean method, and it is symmetrical.<br><br>Like Hausdorff's method, it depends on the distance of the point from the line segment. It calculates the distance of the point from the line segment for all samples of the reference path and all other line segments.<br><br>Hausdorff uses the maximum point-to-path distance, and SSPD uses the mean, which explains why they have almost the same computation time.<br><br>This distance is not time-sensitive and compares the shape and physical distance between two path objects. This method does not require any additional parameters or mapping of different routes. | |

Source: [26]

Distance is the selection criterion for all the distance-based feature selection methods, where both the between and within-class distances are considered.

**Validation methods**

An essential concept in time series research is stationarity, which means that the basic statistics of the time series do not change over time. A series is defined to be stationary [27]. The next step in our methodology is to perform statistical tests to detect statistically significant differences between the reduced databases. For this analysis, a Linear Regression model was chosen as the primary predictive tool. This choice was made due to its simplicity and interpretability. It is worth noting that other regression models can be substituted for the Linear Regression model based on the specific requirements of the analysis. To ensure robust evaluation and mitigate potential overfitting, a 10-fold cross-validation strategy was adopted. This technique partitions the dataset into ten subsets of approximately equal size, ensuring each fold serves as a training and testing set during the evaluation process. The predictive accuracy of the Linear Regression model was evaluated using the Root Mean Square Error (RMSE) and Mean Absolute Error (MAE). RMSE measures the differences between predicted and actual values, while MAE indicates the average magnitude of the errors in a set of predictions. The entire evaluation process was repeated for 10 iterations, with each iteration generating a unique set of RMSE and MAE values. This iterative approach was employed to ensure the reliability and consistency of the evaluation results.

**Result**

In this research, we present the result of the prediction performance of 14 datasets which were selected based on 14 feature selection techniques. Across these 14 techniques, the results of the similarity methods as feature selection methods are provided. We implement seven filter methods, five Wrapper methods, three Embedded methods, and four similarity-based methods. These methods were selected based on Being the most well-known in literature and simply to merely have a comparison between the methods. The Criterion of selected best methods are Root Mean Square Error (RMSE) and Mean Absolute Error (MAE) to show the predictive accuracy of the Linear Regression model. To compare and confidence of efficiency of each dataset chosen based on the FS method, the methods were implemented on the World Bank dataset with different target variables as follows; Adjusted savings consumption of fixed capital, Broad money, Food production index (2014-2016 = 100), Foreign direct investment, net inflows (% of GDP), gdp growth, General government final consumption expenditure (% of GDP), GNI, Gross domestic income, Gross domestic saving, Gross national expenditure (% of GDP), Gross value added at basic prices, Households and NPISHs Final consumption expenditure per capita (constant 2015 US$), Imports of goods and services (constant 2015 US$), Manufacturing, value added (annual % growth), Official exchange rate (LCU per US$, period average), Stocks traded, total value (% of GDP), Total debt service (% of exports of goods, services and primary income), Unemployment, total (% of total labor force) (modeled ILO estimate), Wholesale price index (2010 = 100) and Inflation consumer prices. According to all target values we had 20 different datasets and we used FS methods to choose the best subset of this dataset. Fig 6 shows the result of the evaluation of a 10-fold cross-validation strategy for each FS method. The datasets chosen by these 4 methods had the least Root Mean Square Error (RMSE) and Mean Absolute Error (MAE).

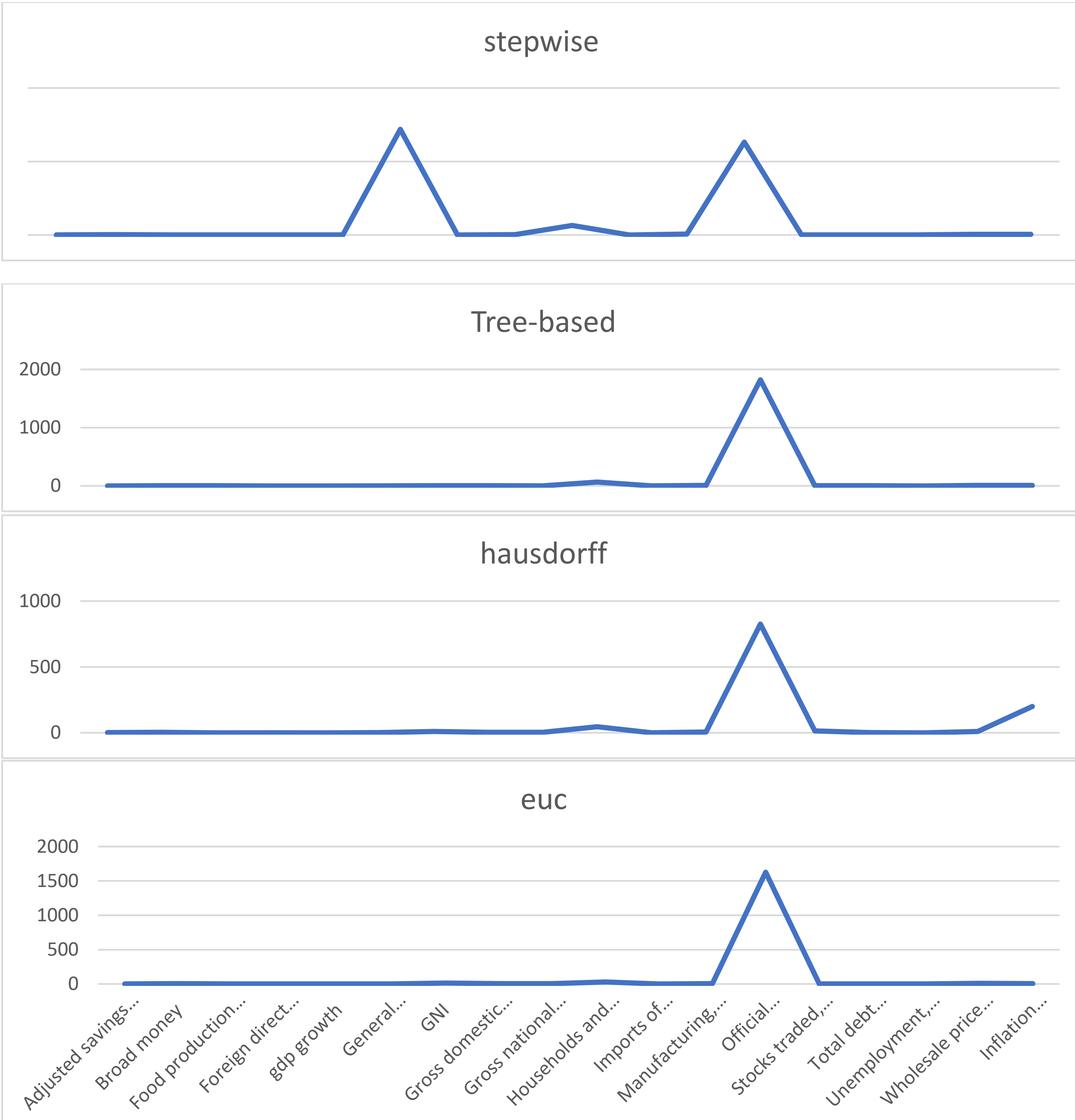

Figure6. Top four feature selection models based on 14 datasets chosen by Feature selection techniques.

Table 4 refers to the result of the average Mean Absolute Error (MAE) of datasets selected by FS methods. The average of MAE based on 20 datasets according to the target value was calculated. the subsets of data in each target value that were selected based on the stepwise FS method had the best prediction performance. The average MAE of the subsets selected based on stepwise had the lowest value. The similarity methods took the next ranking and the subset selected by similarity methods got the lowest MAE average.

| category | methods | average |
|---|---|---|
| Wrappers | stepwise | 32/0299 |
| similarity | frechet | 51/6163 |
| similarity | hausdorff | 62/68829 |
| similarity | sspd | 91/70364 |
| similarity | epr | 91/88632 |
| similarity | dtw | 91/93176 |
| similarity | euc | 95/02939 |
| Embedded | Tree-based | 106/3909 |
| Wrappers | recursive | 270/5572 |
| similarity | lcsso | 292/8808 |
| similarity | edr | 298/4402 |
| Filters | MI_Score | 963/5397 |
| Filters | inf | 1683/06 |
| similarity | Sparse | 3/98E+08 |
| Wrappers | forward | 6/4E+08 |
| Wrappers | simulated_annealing | 8/13E+08 |
| Filters | fisher | 1/83E+09 |
| Embedded | lasso | 3/06E+12 |
| Filters | chi | 4/83E+13 |
| Filters | corrolation | 4/83E+13 |
| Filters | data_dispersion | 8/16E+13 |
| Filters | var | 6/41E+14 |
| Wrappers | backward | 6/47E+14 |

The rank of each FS method to select the best subset of datasets with the lowest MAE Calculated. To get a comprehensive view, the rank of each of the 20 datasets in each method was Averaged. Based on the results, Stepwise Selection, Tree-based, hausdorff, euc, and MI_Score had the best performance, and Recursive Feature Elimination with Cross-Validation and Variance thresholding had a worse performance.

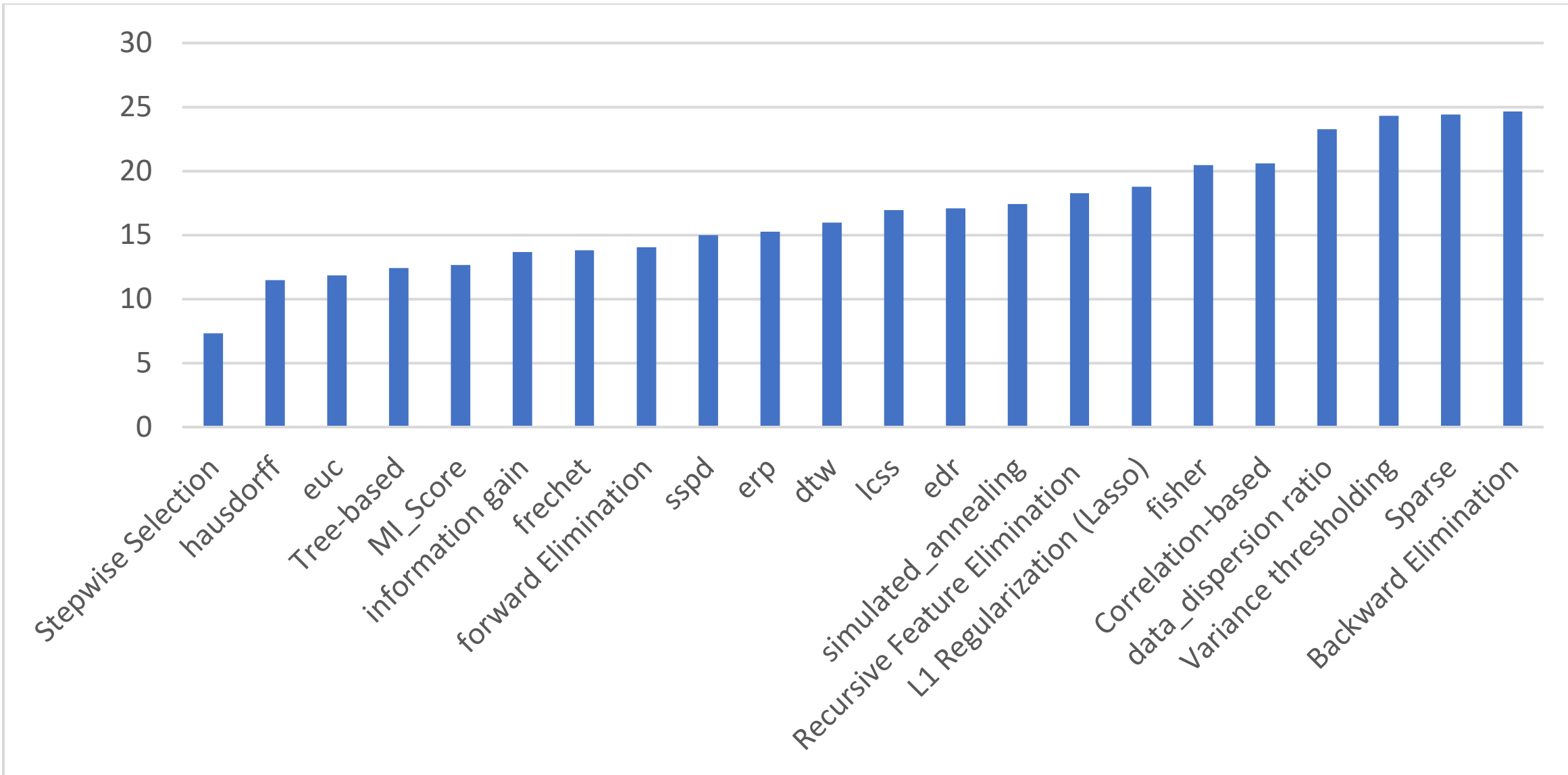

Figure7. the average ranking of MAE selected based on FS methods

The average ranking of the feature selection category shows that on average, similarity methods had performed better than other methods in this ranking. As results show, similarity methods had 9.125 ranks on average.

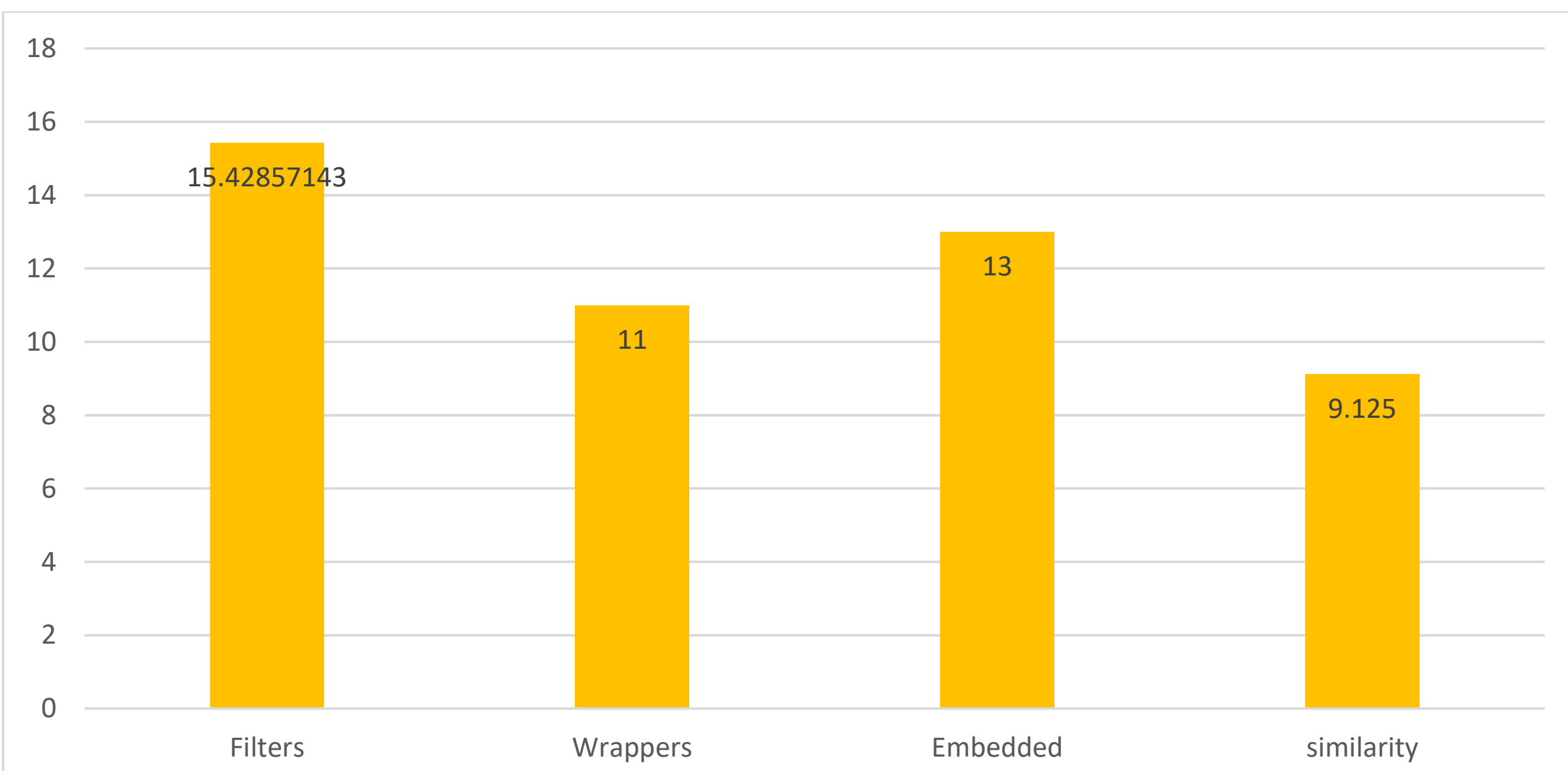

Figure9. The ranking of the category of feature selection methods

## Conclusion

In the current research, we have tested which FS and similarity methods most contribute to improving the performance of models in the forecast of several macroeconomic variables. Analyzed variables include: Adjusted savings consumption of fixed capital, Broad money, Food production index, Imports of goods and services constant 2015 US, Manufacturing value added annual % growth, Official exchange rate LCU per US$ period average, Stocks traded total value, % of GDP, Total debt service % of exports of goods services and primary income, Unemployment total % of total labor force modeled ILO estimate, Wholesale price index 2010 = 100, and Inflation consumer prices. It is in this regard that we have experimented with 23 different ways of feature selection and similarity to understand how to find out the most efficient features that can best predict these macroeconomic indicators.

Time series similarity algorithms are not commonly explored in the literature as feature selection methods. Investigating such algorithms in the current study is to prove their strength by comparing them against traditional FS methods. Every FS and similarity method was applied to each dataset to select good features. Results were evaluated on both MAE and RMSE for each data set. Our results have shown that Stepwise Selection method with Tree-based methods, Hausdorff distance, Euclidean distance, and Mutual Information Score generally performed well compared to other methods, indicating a better predictive accuracy, while Recursive Feature Elimination with Cross-Validation and Variance Thresholding were relatively poorly performed.